\documentclass[a4paper, amsfonts, amssymb, amsmath, prl, showkeys,float,nofootinbib, twocolumn, superscriptaddress]{revtex4-2}
\usepackage[english]{babel}
\usepackage[utf8]{inputenc}
\usepackage[colorinlistoftodos, color=green!40, prependcaption]{todonotes}

\newcommand{\R}{\boldsymbol{\mathsf{R}}}
\usepackage{multirow}
\usepackage{amsthm}
\usepackage{mathtools}
\usepackage{bbold}
\usepackage{physics}
\usepackage{xcolor}
\usepackage{graphicx}
\usepackage[left=13mm,right=13mm,top=35mm,columnsep=15pt]{geometry} 
\usepackage{adjustbox}
\usepackage{placeins}
\usepackage[T1]{fontenc}
\usepackage{csquotes}
\usepackage{natbib}
\usepackage{tikz}
\usetikzlibrary{decorations.markings}
\usetikzlibrary{angles,quotes}

\begin{document}
\title{The Topological Origin of the Peierls-Nabarro Barrier}
\author{Brook J. Hocking}
\affiliation{H. H. Wills Physics Laboratory, University of Bristol, Bristol BS8 1TL, UK}
\author{Helen S.~Ansell}
\affiliation{Department of Physics and Astronomy, University of Pennsylvania,
209 South 33rd Street, Philadelphia, Pennsylvania 19104, USA}
\author{Randall D.~Kamien}
\affiliation{Department of Physics and Astronomy, University of Pennsylvania,
209 South 33rd Street, Philadelphia, Pennsylvania 19104, USA}
\affiliation{Department of Mathematics, University of Pennsylvania,
209 South 33rd Street, Philadelphia, Pennsylvania 19104, USA}
\author{Thomas Machon}
\affiliation{H. H. Wills Physics Laboratory, University of Bristol, Bristol BS8 1TL, UK}

\begin{abstract}
Crystals and other condensed matter systems described by density waves often exhibit dislocations. Here we show, by considering the topology of the ground state manifolds (GSMs) of such systems, that dislocations in the density phase field always split into disclinations, and that the disclinations themselves are constrained to sit at particular points in the GSM. Consequently, the topology of the GSM forbids zero-energy dislocation glide, giving rise to a Peirels-Nabarro barrier.
\end{abstract}

\maketitle
The symmetries of crystals are encoded by groups: the space group of each possible crystalline arrangement is composed of discrete translations, rotations, glides, mirrors, and screws that form finite subgroups of the Euclidean group in the dimension of interest.  As a result, and as strikingly measured in X-ray diffraction, physical crystals provide a near Platonic manifestation of rigid geometry via their periodicities.   These symmetries are encoded in the ground state manifold (GSM) that parameterizes degenerate ground states.  The coordinates of the GSM are the Nambu-Goldstone modes of the system -- the zero-energy deformations.  When the GSM is spanned by these modes (moduli), homotopy theory allows for the classification of topological defects \cite{mermin79,pollard,kleman78}.  For instance, the discrete translational symmetry of the crystal leads to dislocations with ``charges'' that are integer combinations of the lattice basis vectors -- the Burgers vector.   In the classic discussion by Peierls, the dislocations move in the background of the lattice and it follows that an energy cost associated with dislocation glide inherits the crystal periodicity \cite{peierls40,nabarro47}.  The ``Nabarro-Peierls'' barrier is the keystone upon which dislocation motion is based. 

 What happens, however, when the crystal is very soft?  As a concrete example, recall the Tonks-Girardeau \cite{tonks36,girardeau60} gas of $N$ impenatrable rods, length $\ell$ on a line $L$ long.   Because of the excluded length interaction the particles are distinguishable since their order on the line is fixed: the finite-energy configuration space breaks up into $N!$ disconnected sectors, each labelled by the list of nearest neighbors --  the topology of the rod configuration space.
As two adjacent rods swap positions we are moved from one sector of phase space to another, passing through a forbidden, infinite energy configuration corresponding to a change in topology.  In this letter, we will make this connection precise and show that the barrier to dislocation glide arises from topological considerations and that, for instance, Peierls' form of the energy barrier is a consequence of a transition between topological sectors.  It is essential to fully explore the GSM and to include the point group symmetries of the crystal in order to introduce disclinations along with the dislocations.  As noted in \cite{mermin79,poenaru81}, homotopy theory fails to characterize these defects because the Nambu-Goldstone moduli do not span the GSM \cite{low02}.  As discussed in \cite{chen09,pevnyi14,machon19} this can be resolved via the use of Morse-theoretic tools. Specifically, between two maxima in the density wave, reflected in X-ray diffraction, there must be saddles and minima.  These critical points, on the same lengthscale as the Peierls potential, are the essence of the topological barrier.

\begin{figure}[!t]
    \centering
    \includegraphics[width=0.5\textwidth]{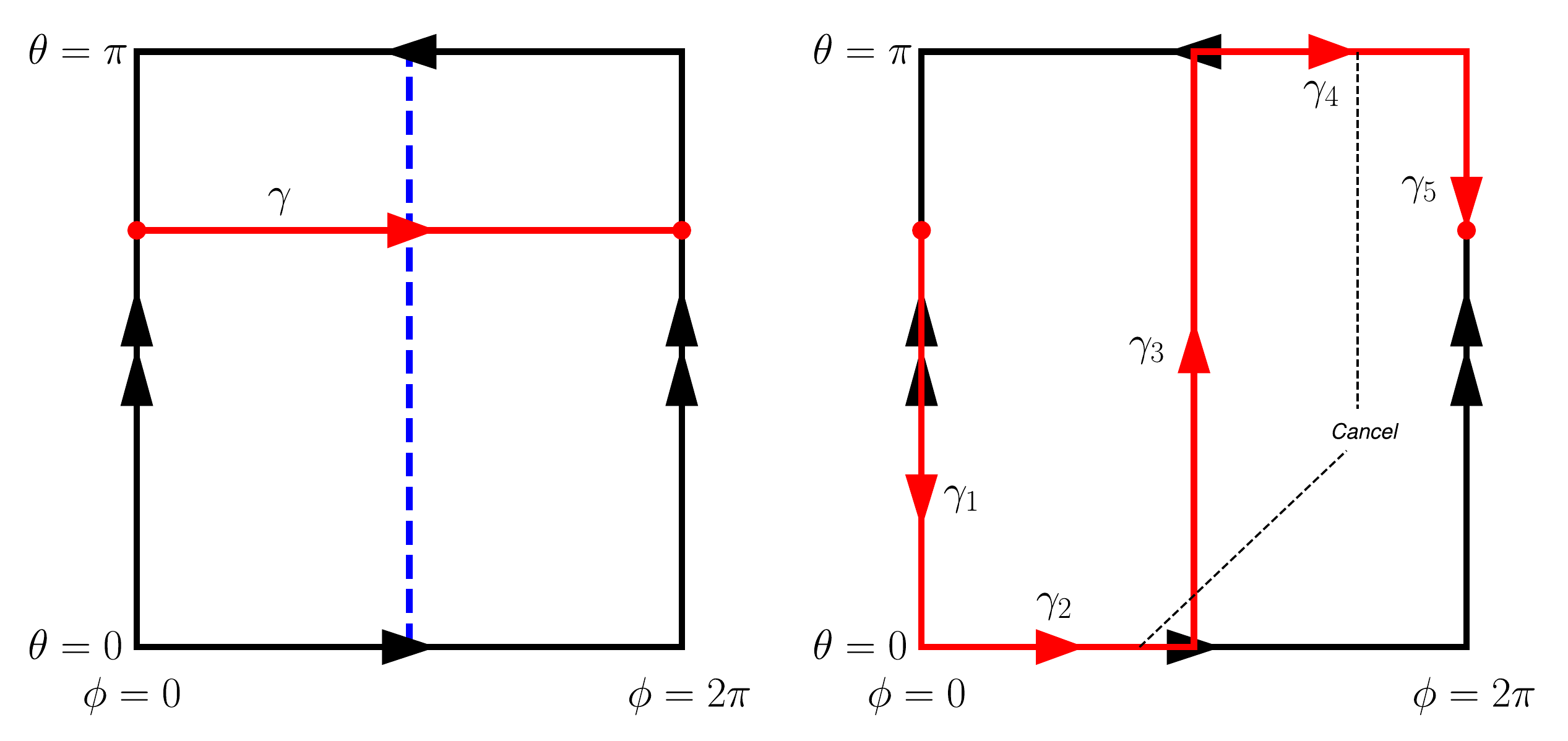}
    \caption{Left: The two-dimensional smectic GSM (Klein bottle) featuring a path corresponding to a dislocation (red) winding in $\phi$. The dashed blue line is at $\phi = \pi$ and along with the vertical at $\phi = 0$ comprise the allowed disclination locations. Right: Decomposition of dislocations into disclinations in the GSM. The red path is deformed, employing the equivalence relations for the Klein bottle  the segments $\gamma_2$ and $\gamma_4$ are now identical but with opposite orientation. The remaining segments $\gamma_1$ and  $\gamma_3$ correspond to disclinations at a density maximum ($\phi=0)$ and minimum ($\phi = \pi$) respectively -- the dislocation has been decomposed into a pair of disclinations as shown in the upper left of Fig. 2.  This defect complexion does not require $\delta\rho$ to vanish anywhere.}
    \label{fig:smecticgsm}
\end{figure}

\begin{figure}[t]
    \centering
    \includegraphics[scale=0.4]{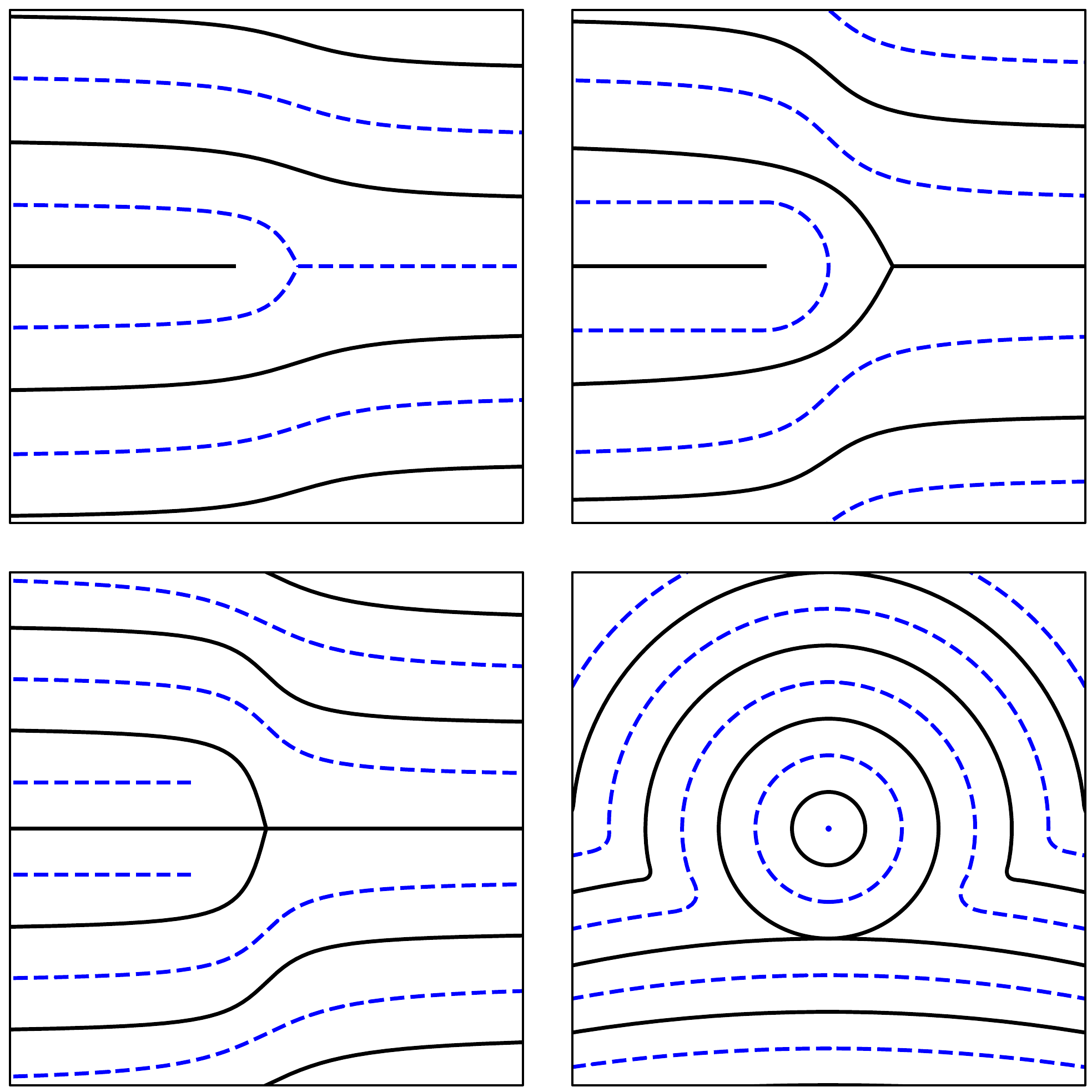}
    \caption{Configurations of dislocations in terms of defects on phase maxima (solid black) and minima (dashed blue).  Clockwise from upper left: 1) A standard charge 1 dislocation that adds one layer.  It has a $\pm$ pair of $\pi$ phase disclinations on density minima/maxima; 2) A charge 2 dislocation with both disclinations ($\pm\pi$) lying on density maxima; 3) A charge $+2\pi$ disclination cannot connect to other layers and can only participate in a {\sl pincement}; 4) A charge 2 dislocation made of a $-2\pi$ disclination and two $+\pi$ disclinations.}
    \label{fig:smecticconf}
\end{figure}

We first consider the two-dimensional smectic which captures the essence of this work. The layer structure of the smectic ground state can be described by the phase field $\Phi = \mathbf{k} \!\cdot\! \mathbf{x} + \phi$. The density is reconstructed via $ \rho = \rho_0 + \delta \rho \cos \Phi$, the layers sit on density maxima ($\cos \Phi =1$).
The space of ground states is described by two variables: the phase at the origin $\phi$ with $\phi\cong\phi + 2\pi \mathbb{Z}$, reflecting the periodic nature of the ground state, and the layer normal $\mathbf{n} =  \nabla \Phi / | \nabla \Phi |$ with $\mathbf{n} \cong -\mathbf{n}$, reflecting the unoriented nature of the smectic and parallel (or antiparallel) to $\mathbf{k}$.
In two dimensions, the director can be specified by an angle $\theta$; since the layers are unoriented, however, rotating the crystal by $\pi$ about the origin is equivalent to no rotation but with a shift in $\phi$. We then arrive at the \textit{ground state manifold} (GSM) described by $(\theta,\phi)$ with identification $(\theta ,\phi) \cong (\theta+ \pi,-\phi)$ -- the GSM is the Klein bottle \cite{trebin82}, shown in Fig.~\ref{fig:smecticgsm} (left).  When $\phi$ is not defined, $\delta\rho$ must vanish resulting in an energy ``barrier" associated with the melted state.

\begin{figure*}[t]
    \centering
    \includegraphics[scale=1]{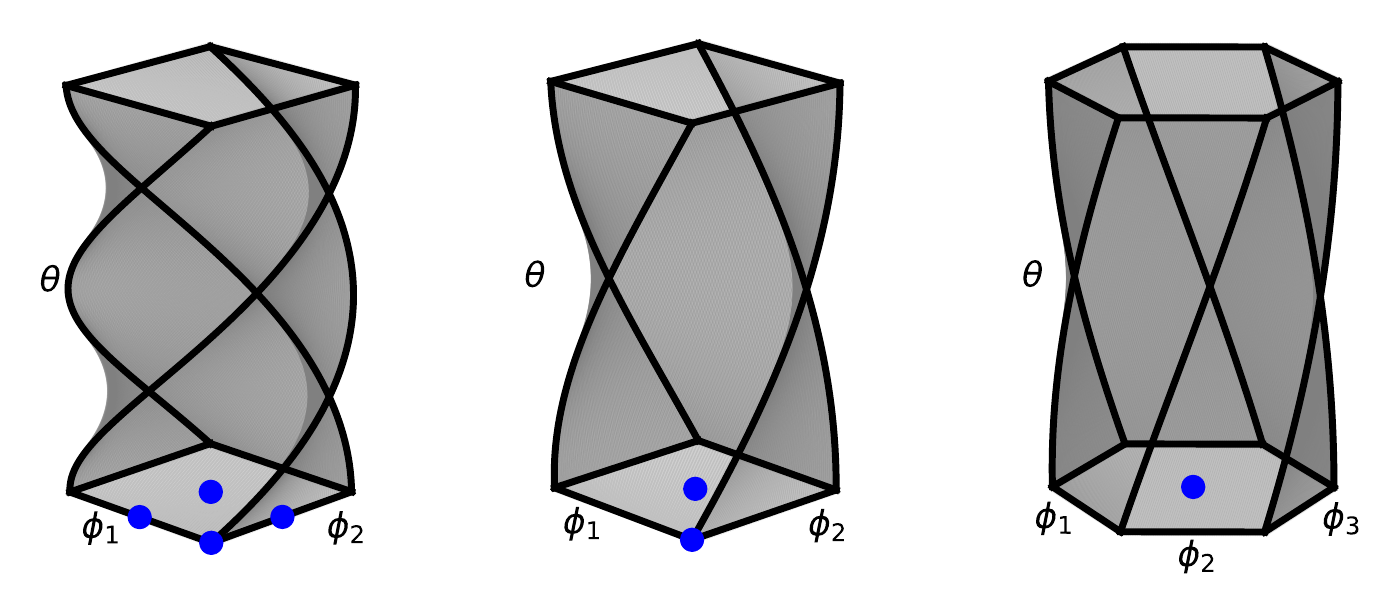}
    \caption{The GSMs for $\pi$, $\pi/2$, and $\pi/3$ rotationally symmetric lattices respectively. The top is glued directly to the bottom without any further twisting.  Dots indicate allowed phases for the weakest strength disclinations.}
    \label{fig:gsmdislocations}
\end{figure*}
\begin{table*}
\squeezetable
\begin{ruledtabular}
\begin{tabular}{c c c c c} 
 
 Lattices & $\theta_0$ & GSM $(\theta,\phi_1,\phi_2) \cong$ & Disclination Angles & GSM Locations \\ [0.5ex] 
 \hline
 
 O, R, CR & $\pi$ &  $(\theta + \pi, -\phi_1,-\phi_2)$ & $\pm \pi$ & $(0,0) \enspace (\pi,0) \enspace (0,\pi) \enspace (\pi,\pi)$ \\
 \hline
  \multirow{2}{*} {S} & \multirow{2}{*} {$\pi/2$} & \multirow{2}{*} {$(\theta + \pi/2, \phi_2,-\phi_1)$} &  $\pm \pi/2$ & $(0,0) \enspace (\pi,\pi)$ \\ & & & $\pm \pi$ & $(0,0) \enspace (\pi,0) \enspace (0,\pi) \enspace (\pi,\pi)$ \\
 \hline
 \multirow{3}{*} {T} & \multirow{3}{*} {$\pi/3$} & \multirow{3}{*} {$(\theta + \pi/3, \phi_1 + \phi_2,-\phi_1)$} & $\pm \pi/3$ & $(0,0)$ \\ & & & $\pm 2\pi/3$ & $(0,0) \enspace (2\pi/3, 2\pi/3) \enspace (4\pi/3,4\pi/3)$ \\ & & & $\pm \pi$ & $ (0,0) \enspace (\pi,0) \enspace (0,\pi) \enspace (\pi,\pi)$ \\
 
\end{tabular}
\end{ruledtabular}
\caption{Groundstate manifolds (GSMs), disclinations and their GSM locations for the oblique (O), rectangular (R), centered rectangular (CR), square (S), triangular (T).}
\label{tab:summary}
\end{table*}

Every regular point in the smectic can be associated to a point in the GSM, while points that do not allow this are defects. Defects in the smectic are of two types: disclinations (singularities in the orientation variable $\theta$) and dislocations  (singularities in the density variable $\phi$).
First, consider a $\pm \pi$ disclination: while the orientation ${\bf n}$ is undefined there, the density field $\cos \Phi$ may remain continuous. A measuring path in the sample around the disclination maps to a closed loop in the GSM that identifies $(0,\phi)$ with $(\pi, - \phi)$. Though the measuring loop can never be shrunk to a point, if we were to contract the measuring path in the sample, $\phi$ must approach a constant limiting value in the GSM. But the condition on the GSM $(0,\phi) \cong (\pi, - \phi)$ implies that this constant value of $\phi$ must satisfy $\phi \cong - \phi$, so $\phi = 0$ or $\pi$. Hence $\pm \pi$ disclinations in smectics must sit on density minima/maxima~\cite{chen09, machon19}.  We can also see that only these values of $\phi$ allow a pure disclination -- other values of $\phi$ would prevent the loop from closing in the GSM.
Recall that the smectic groundstate has a single Goldstone mode, arising from the zero-energy deformation $\phi \mapsto \phi + \epsilon$, indeed this deformation is a zero energy translation for {\em any} defect-free smectic configuration and amounts to shifting the origin. However, in the presence of disclinations it is no longer possible to make the transformation $\phi \mapsto \phi + \epsilon$, since the value of $\phi$ on disclinations is topologically fixed. In other words, the presence of disclinations implies the smectic energy no longer has a ${\rm U}(1)$ symmetry. In particular, de Gennes' model of the smectic~\cite{degennes72} cannot accurately describe disclinations~\cite{pevnyi14,xia21}.

How does this affect dislocations? Dislocations are necessarily made of a topologically protected arrangement of disclinations with zero net charge. This follows because lines of phase maxima must be separated by lines of phase minima \cite{machon19}.  In the GSM a charge one dislocation is represented by a path winding $2 \pi$ in $\phi$ (Fig.~\ref{fig:smecticgsm}, left), by manipulating the path, using the topology of the Klein bottle, it is converted in to two disclinations, one on a density maximum and one on a density minimum (Fig.~\ref{fig:smecticgsm}, right); the real-space realisation is shown in Fig. \ref{fig:smecticconf} (top left).  By including more maxima/minima lines, we can increase the charge by changing the ``length'' of the disclination dipole.  Alternatively, one could, for instance, attempt to construct a charge two dislocation out of a $\pm 2\pi$ disclination pair.  However, because $+2\pi$ disclinations cannot connect to distant layers, it is necessary to balance the $-2\pi$ disclination with {\sl two} $\pi$ disclinations.  As there are no disclinations with charge higher than $2\pi$ \cite{poenaru81} it follows that {\sl all} dislocations must contain $\pi$ disclinations on either a phase maximum or minimum.  Thus, purely on symmetry grounds, a dislocation glide is forbidden by the topology of the GSM: the Peierls-Nabarro barrier.  Indeed, in order for a dislocation to glide, the smectic order itself must melt, allowing the disclinations to hop from one layer to another, moving from one topological sector to another. 

We now extend this discussion to the five Bravais lattices describing two-dimensional crystals, beginning with a description of the GSM. A two-dimensional lattice is determined by two basis vectors ${\bf e}_1$ and ${\bf e}_2$ with reciprocal lattice vectors ${\bf k}_1$ and ${\bf k}_2$, ${\bf k}_i\cdot{\bf e}_j=\delta_{ij}$. In a two-dimensional lattice, the vertices can be specified by maxima of $\cos \Phi_1+ \cos \Phi_2$ where the two phase fields are $\Phi_i = \mathbf{k}_i\! \cdot\! \mathbf{x} + \phi_{i}$ ($i=1,2$).  Reusing $\theta$ to specify the orientation of the lattice with respect to a fixed-axis, the GSM is parameterized by $(\theta, \phi_1, \phi_2)$. To study the translational symmetry we write
\begin{equation}
\Phi_i = \mathbf{k}_i \cdot \mathbf{x} + \phi_{i} = {\bf k}_i \cdot ({\bf x}+ {\bf x}_0),
\end{equation}
where ${\bf x}_0$ is (minus) the position vector of the lattice point corresponding to $\Phi_1 = \Phi_2 = 0$. A translation by any lattice vector is a symmetry of the lattice, so
${\bf x}_0 \cong {\bf x}_0 + {\bf e}_1\mathbb{Z}+{\bf e}_2\mathbb{Z}$. This symmetry allows us to choose ${\bf x}_0$ in the unit cell. Now, suppose that a rotation by $\theta_0$, denoted $\R$, about $-{\bf x}_0$ is a symmetry of the lattice. Applying the rotation gives
\begin{align}
    \Phi_i^{'} & = (\R\mathbf{k}_i) \cdot (\mathbf{x}+{\bf x}_0),
\end{align}
which is equivalent to a rotation about the origin by $\bf R$ and a translation of $\phi_i$ by $[(\mathbb{1}-\R) \mathbf{k}_i]\cdot {\bf x}_0$,
\begin{align}
    \Phi_i^{'} & = (\R\mathbf{k}_i) \cdot (\mathbf{x}+ \R{\bf x}_0) -[(\mathbb{1}-\R) \mathbf{k}_i] \cdot {\bf x}_0.
\end{align}
Since we know $\Phi_i \cong \Phi_i^{'}$, we find the equivalence relation
\begin{equation}(\theta, \phi_1, \phi_2) \cong\bigg(\theta+\theta_0, (\R{\bf k}_1) \cdot {\bf x}_0, (\R {\bf k}_2) \cdot {\bf x}_0 \bigg),
\end{equation}
with $\phi_{i} = {\bf k}_i \cdot {\bf x}_0$. Along with the translational symmetry, this specifies the GSM for the two-dimensional Bravais lattices. The basic rotational symmetries of the five Bravais lattices are $\theta_0=\pi$ for the oblique (O), rectangular (R) and centered rectangular (CR), $\pi/2$ for the square (S), and $\pi/3$ for the triangular (T), in each case we obtain a topologically distinct GSM, as shown in Fig.~\ref{fig:gsmdislocations} and listed in Tab.~\ref{tab:summary}. These can be described topologically as the product of a unit cell (a torus) with the interval $[0, \theta_0]$, such that the $0$ and $\theta_0$ unit cells are glued together with a $\theta_0$ twist.  Formally, this defines an element of ${\rm SL}(2, \mathbb{Z})$, a new choice of basis vectors, for the closed path from $0$ to $\theta_0$.  The matrices for $\theta_0 = \pi$, $\pi/2$, $\pi/3$ are of finite orders 2, 4, and 6, corresponding to the allowed point groups of a lattice without basis.
Finally, It should be noted that although the function $\cos \Phi_1+ \cos \Phi_2$ is sufficient to describe the symmetries of the lattice via maxima points, it is not guaranteed that the density field has the same symmetry.  This is only an issue in the triangular lattice.  In order to have the density enjoy the full, six-fold symmetry around each lattice point we must assign a third phase field $\Phi_3 \cong - \Phi_1 - \Phi_2$ (for the convention of $\mathbf{k_1}, \mathbf{k_2}$ differing by angle $2\pi/3$), and promote the density function to $\rho-\rho_0\propto(\cos \Phi_1+ \cos \Phi_2 + \cos \Phi_3)$.  Note that if any two of $\Phi_1,\Phi_2$, or $\Phi_3$ are integer multiples of $2\pi$ then so is the third -- the density maxima are unchanged by this necessary embellishment to the density wave.

How do crystal dislocations appear when viewed in terms of the phase fields? Since the lattice points are defined to sit where $\cos \Phi_1$ and $\cos \Phi_2$ are both maximum, and individually each phase field $\Phi_i = \mathbf{k}_i\! \cdot\! \mathbf{x} + \phi_{i}$ describes a smectic configuration, it can be seen that a defect-free two-dimensional lattice is the intersection of two smectics.  However, the rotational symmetries mix the $\Phi_i$ and can impose further constraints on the locations of the phase disclinations. 

Now consider a dislocation in a crystal lattice. The density field around a dislocation must remain non-singular, hence any map from the sample into the GSM in the vicinity of the dislocation must contain singularities only in the orientation of the phase fields -- these are phase field disclinations.  It is important here that we are focussing on disclinations of the phase fields and {\sl not} the traditional disclinations in the crystal.  Sometimes they coincide but, as we will see, not always.  For instance, in a square lattice a dislocation breaks up into two phase disclinations but, on the crystal itself there is only one disclination with its negative partner on the dual lattice. On a rectangular lattice, the dislocation remains localised to a unit cell, in terms of the phase field however, it naturally splits into two phase disclinations in $\phi_1$ while $\phi_2$ remains smooth, the analogy to the smectic is clear. Note that while, in principle, $\pi$ disclinations may appear as isolated defects in the O, R, and CR lattices; the individual disclinations are not typically seen due the large energetic cost.
\begin{figure}[!t]
    \centering
    \includegraphics[scale=0.28]{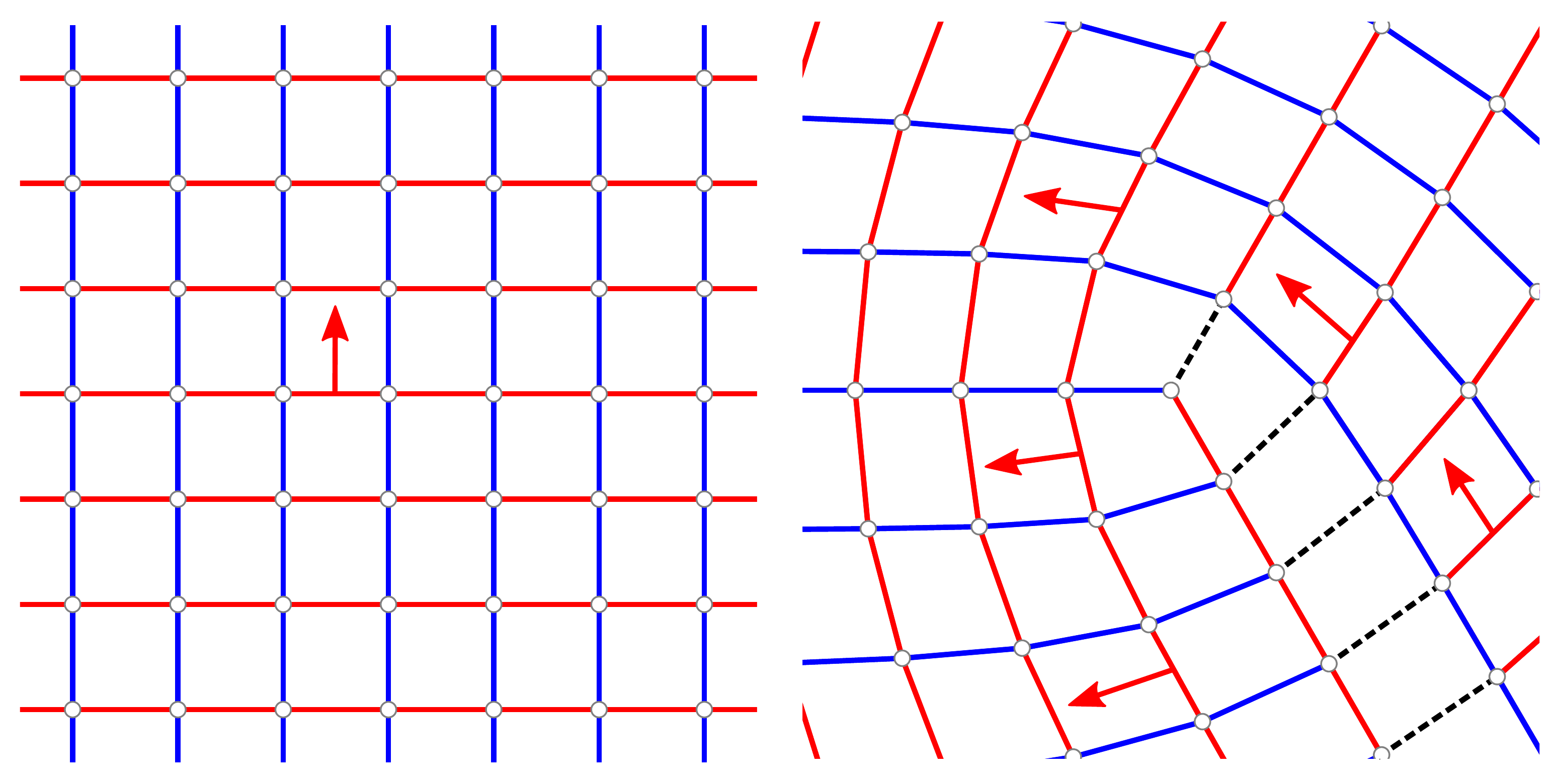}
    \caption{Square Lattice with maxima lines being attributed to $\Phi_1$ (red) or $\Phi_2$ (blue). Arrows indicate the normal vector with convention $\mathbf{n} \!\parallel\! \nabla \Phi_1$. In the presence of a disclination, a complete and consistent assignment of $\Phi_1$ or $\Phi_2$ cannot be achieved, as indicated by the unassigned sections. }
    \label{fig:sqaurelattices}
\end{figure}

For the square and triangular lattices the situation is more complex. In these cases the rotations by $\pi/2$ and $\pi/3$ transform the phase fields among themselves; so that in the vicinity of a defect there can be no consistent labelling of $\Phi_1$ and $\Phi_2$. Despite this, dislocations still split into phase disclinations as before, as shown in Fig.~\ref{fig:sqaurelattices}. Despite the seemingly ambiguous nature of the phases around a disclination, when a measuring path is mapped to the GSM no such discontinuity is seen, since the GSM is the space of states modulo equivalent states; disclinations $\{d_i\}$ in a lattice $\Omega$ correspond to a \textit{smooth} map $f:\Omega /\{d_i\} \to GSM$. 
In a manner directly analogous to the smectic, the topology of the GSM dictates the locations of phase disclinations. Consider shrinking the radius of a measuring loop around a disclination, such that $\phi_1$ and $\phi_2$ approach constant values, but $\theta \to \theta + n\theta_0$. This can then only occur at locations in the GSM where $(\phi_1,\phi_2)$ is invariant under $\theta \to \theta + n\theta_0$, so that $\exists\; a,b\in\mathbb{Z}$ such that
\begin{equation}
    \R_{\pm \theta_0} \begin{pmatrix} \phi_1 \\ \phi_2 \\ \end{pmatrix} = \begin{pmatrix} \phi_1 \\ \phi_2 \\ \end{pmatrix} + \begin{pmatrix} 2\pi a \\ 2\pi b \\ \end{pmatrix}
\end{equation}
Consider the square lattice as an example, we have seen the twisting of the GSM is described by $(\theta,\phi_1,\phi_2) \cong (\theta + \pi/2,\phi_2,-\phi_1)$. This has solutions when $\phi_1$ and $\phi_2$ are both minima or both maxima: the disclinations are confined to one of two positions within the unit cell, here corresponding to crystal disclinations lying on vertices of the lattice or dual lattice. The full list of allowed phase disclination densities is given in Tab.~\ref{tab:summary}.  There we see that the standard ``5'' and ``7'' disclinations of the triangular lattice always lie on double maxima -- in agreement with the classic theory of crystal disclinations employed in two-dimensional melting \cite{halperinnelson78,halperinnelson79,young79}.  We note that in the square and triangular lattices, disclinations of higher winding than the elementary $\pi/2$ or $\pi/3$ may exist -- the lower symmetry gives such disclinations greater freedom of location. 

Since phase disclinations must lie on a restricted set of points in the unit cell in order for the phase field to be continuous, it follows that there is a Peierls-Nabarro barrier for dislocations that must contain disclinations; the ground state symmetry immobilises the underlying disclination structure of a lattice dislocation. A dislocation cannot glide without first melting the density wave order.  This is akin to closing the mass gap in topological insulators: in order to transition from one topological sector to the other the 
density wave must vanish, destroying the ability to define the phase fields $\Phi_i$ and allowing topological invariants to ``tunnel'' from one value to another \cite{TKNN82}.

We have shown that in two-dimensions the Peierls-Nabarro barrier follows from topological considerations.  What happens in three-dimensions?  In this case disclinations and dislocations are described by lines or closed loops in the sample, not points.  In cross section, perpendicular to the defect lines, our arguments continue to hold and there remains a topological Peierls-Nabarro barrier to glide.  However, when the defect lines bend, there are likely to be further, global, topological constraints as there are in, for instance, biaxial nematics \cite{mermin79}, cholesterics \cite{kleman69,beller14}, and smectics \cite{mosna16,machon19}.  Whether these additional constraints can be resolved or lead to additional topological invariants remains an open question.

We dedicate this paper to the memory of Maurice Kleman,  acknowledge the fine and disruptive commentary of J.H. Hannay, and thank G. Grason and N. Spaldin for useful discussion. H.S.A. and R.D.K. were supported by NSF MRSEC Grant DMR-1720530 and a Simons Investigator Grant from the Simons Foundation to R.D.K. B.J.H  and T.M were supported by EPSRC grant EP/T517872/1.

\end{document}